\documentclass[12pt]{iopart}

\usepackage[utf8]{inputenc}

\usepackage{graphicx}
\usepackage{iopams}
\usepackage{amsmath, amsthm, amssymb, amsfonts, amstext}
\usepackage{hyperref}
\hypersetup{
    colorlinks=true,
    linkcolor=blue,
    filecolor=magenta,      
    urlcolor=cyan,
}
\begin{document}

\title{Equations of Motion Formulation of a Pendulum Containing N-point Masses}

\author{Boran Yeşilyurt}

\address{Orta Doğu Teknik Üniversitesi, Üniversiteler Mahallesi, Dumlupınar Bulvarı No:1 06800 Çankaya Ankara/TÜRKİYE}
\ead{yesilyurt.boran@metu.edu.tr}
\vspace{10pt}
\begin{indented}
\item[]
\end{indented}

\begin{abstract}
This paper presents a general formulation of equations of motion of a pendulum with n point mass by use of two different methods. The first one is obtained by using Lagrange Mechanics and mathematical induction(inspection), and the second one is derived by defining a vector. Today, these equations can be obtained by employing numerous programs; however, this study gives a very compact form of these equations that is more efficient than solving Euler-Lagrange Equations for every pendulum with more complex structures than simple or double pendulum. Additionally, we investigate what will happen to our n-point mass system when we take limit as number of point masses goes infinity under well-defined assumptions. We find out that it converges to hanging rope system.
\end{abstract}
\vspace{2pc}
\noindent{\it Keywords}: Pendulum, Classical Mechanics
\section{Introduction}
Even though it is not a commonly encountered problem in numerous areas of Physics, double and triple pendulums are examined in the study of chaos and classical mechanics. However, equations of motion of these systems obtained by Lagrange Mechanics can be long and complicated. It is evident that these equations may be acquired with the help of computers today, but the way of obtaining these equations by computers involves taking partial derivatives, which may cause some trouble for higher systems containing more than three point masses. Thus, this new formulation of equations of motion of n-point mass pendulum systems might quicken this process. By doing so, it may help the study of chaos in these particular systems. The pendulum systems which we will investigate in this paper consists of point masses and movable joints. After we obtain the general formula, we will examine the small oscillations.

\section{First Way of Derivation}
\begin{figure}[t]
\includegraphics[scale=0.75]{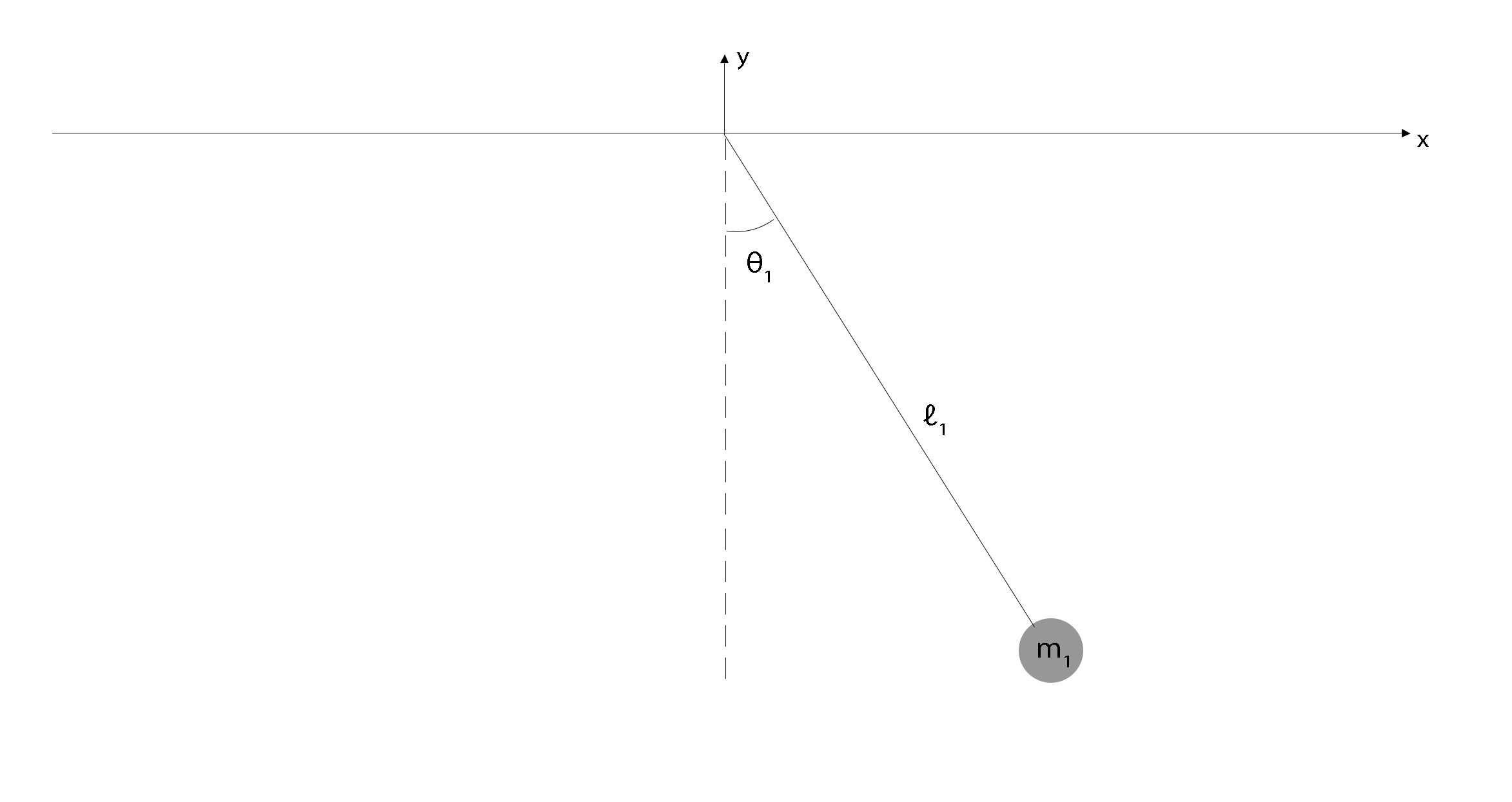}
\centering 
	\caption{A Simple Pendulum} 
	\label{fig:1} 
\end{figure}

\begin{figure}
\centering 
	\includegraphics[scale=0.6]{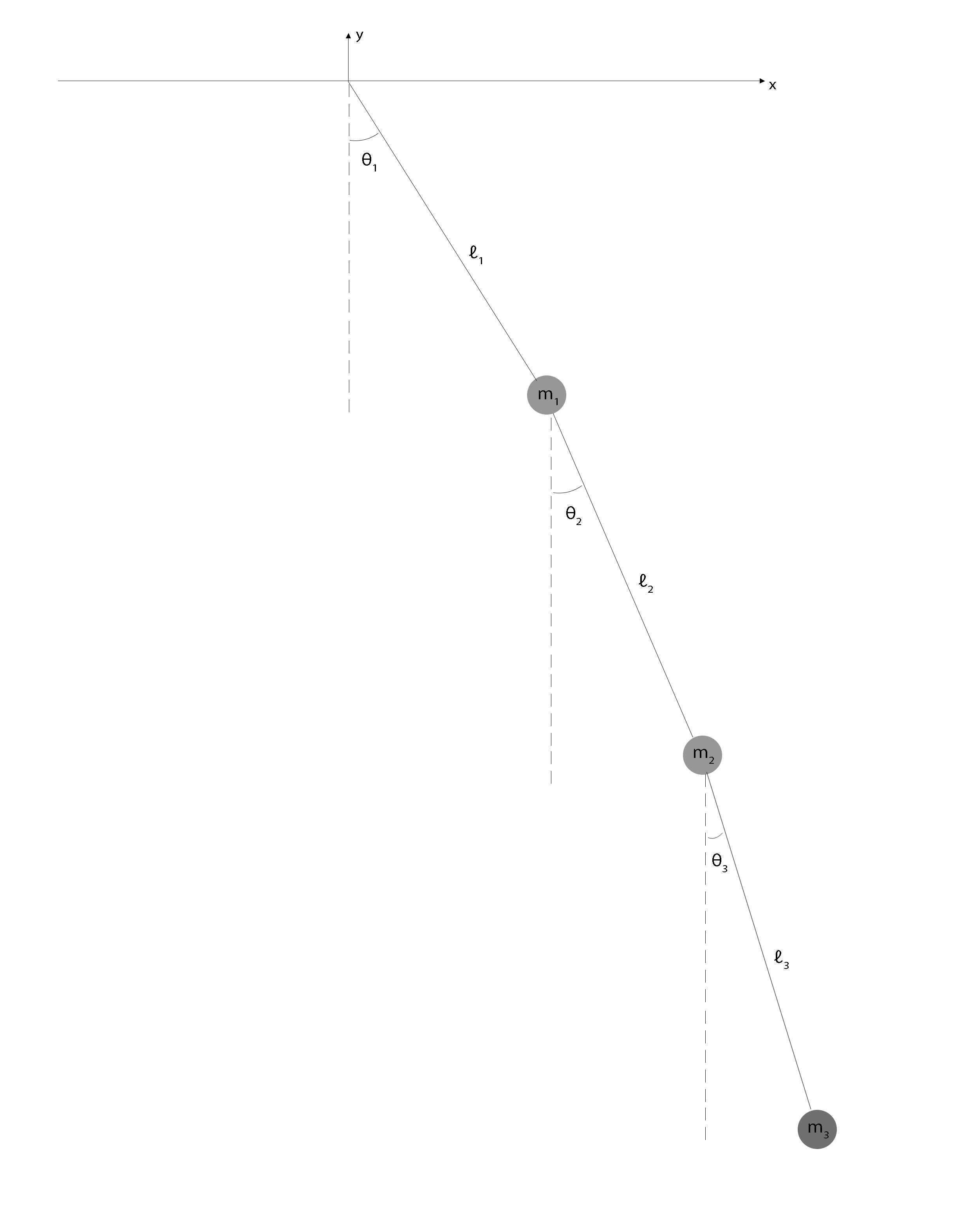}
	\caption{A Triple Pendulum}
	\label{fig:3}
\end{figure}

The derivation of the final form of the formula will be mainly based on induction. Let us start with the Lagrangian and equations of motion of simple pendulum and double pendulum.
According to \ref{fig:1}
\begin{equation}
\ddot{\theta}+\frac{g}{l}\sin(\theta)=0
\label{1}
\end{equation} 
\begin{equation}
L=T-U=\frac{1}{2}ml^2\dot{\theta}^2+mgl\cos(\theta)\quad\quad 
\label{2}
\end{equation} 
These are the aforementioned equations for simple pendulum\cite{la}. 
\begin{figure}[b]
\centering 
	\includegraphics[scale=0.65]{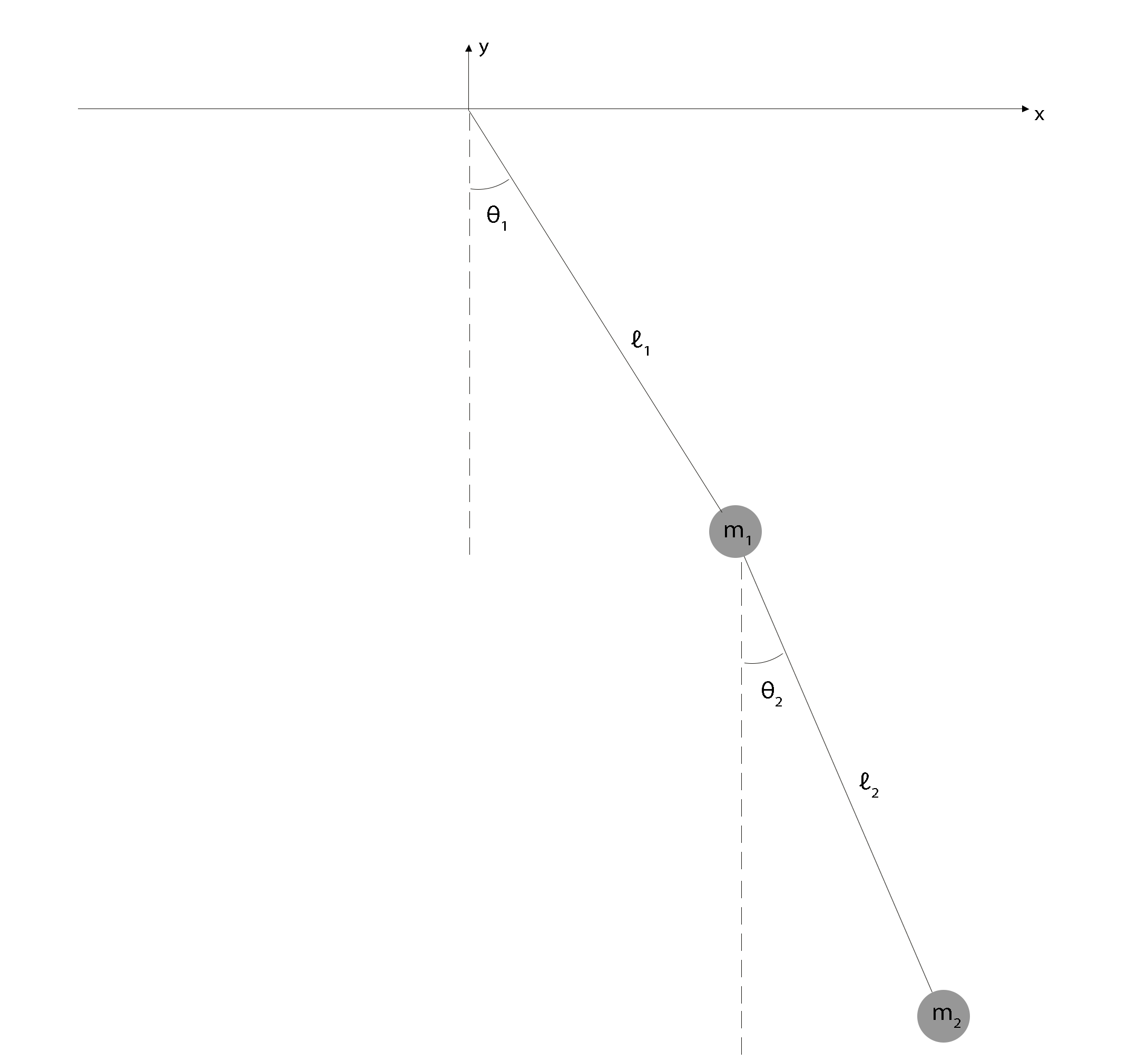}
	\caption{A Double Pendulum}
	\label{fig:2}
\end{figure} 
Before we give the equations for double pendulum \ref{fig:2}, we need to clarify some points. \\
\begin{equation}
x_1=l_1\sin(\theta_1)
\end{equation}
\begin{equation}
y_1=-l_1\cos(\theta_1)
\end{equation}
\begin{equation}
x_2=l_1\sin(\theta_1)+ l_2\sin(\theta_2)
\end{equation}
\begin{equation}
y_2=-(l_1\cos(\theta_1)+ l_2\cos(\theta_2))
\end{equation}
Above four formulas give the positions of point masses. Then, for the first point mass 
\begin{equation}
dx_1=l_1\cos(\theta_1)d\theta_1\quad and \quad dy_1=l_1sin(\theta_1)d\theta_1
\end{equation}
Thus, 
\begin{equation}
(dx_1/dt)^2+(dy_1/dt)^2=l_1^2\dot{\theta_1}^2
\end{equation} 
Hence 
\begin{equation}
K_1=\frac{1}{2}m_1l_1^2\dot{\theta_1}^2
\end{equation}
Also, 
\begin{equation}
U_1=-m_1gl_1\cos(\theta_1)
\end{equation} 
Consequently,\begin{equation}
L_1=K_1-U_1=\frac{1}{2}m_1l_1^2\dot{\theta_1}^2+m_1gl_1\cos(\theta_1).
\label{3}
\end{equation} 
Now, \begin{equation}
dx_2=l_1\cos(\theta_1)d\theta_1+l_2\cos(\theta_2)d\theta_2
\label{4}
\end{equation}
\begin{equation}
dy_2=l_1\sin(\theta_1)d\theta_1+l_2\sin(\theta_2)d\theta_2
\label{5}
\end{equation}
Then,\begin{equation}
\quad (dx_2/dt)^2+(dy_2/dt)^2= l_1^2\dot{\theta_1}+l_2^2\dot{\theta_2}^2+2l_1l_2\cos(\theta_1-\theta_2)\dot{\theta_1}\dot{\theta_2}
\label{6}
\end{equation}
Hence,
\begin{equation}
K_2=\frac{1}{2}m_2(l_1^2\dot{\theta_1}+l_2^2\dot{\theta_2}^2+l_1l_2\cos(\theta_1-\theta_2)\dot{\theta_1}\dot{\theta_2})
\end{equation}   
\begin{equation}
 U_2=-m_2g(l_1\cos(\theta_1)+ l_2\cos(\theta_2))
\end{equation}
Consequently, \begin{equation}
 L_2=K_2-U_2=\frac{1}{2}m_2(l_1^2\dot{\theta_1}+l_2^2\dot{\theta_2}^2+2l_1l_2\cos(\theta_1-\theta_2)\dot{\theta_1}\dot{\theta_2})+m_2g(l_1\cos(\theta_1)+ l_2\cos(\theta_2))
 \end{equation}
Then, the lagrangian of double pendulum is \begin{equation}
\begin{aligned}
L &=L_1+L_2=
\frac{1}{2}m_1l_1^2\dot{\theta_1}^2+m_1gl_1\cos(\theta_1) \\
&+\frac{1}{2}m_2(l_1^2\dot{\theta_1} +l_2^2\dot{\theta_2}^2+2l_1l_2\cos(\theta_1-\theta_2)\dot{\theta_1}\dot{\theta_2})
+m_2g(l_1\cos(\theta_1)+ l_2\cos(\theta_2))
\end{aligned}
\label{7}
\end{equation}
Then by solving Euler-Lagrange Equations ($\frac{d}{dt}(\frac{\partial L}{\partial\dot{q}})-\frac{\partial L}{\partial q}=0$) for $\theta_1$, we get \begin{equation}
(m_1+m_2)l_1^2\ddot{\theta_1}+m_2l_1l_2\ddot{\theta_2}cos(\theta_1-\theta_2)+m_2l_1l_2\dot{\theta_2}^2sin(\theta_1-\theta_2)+(m_1+m_2)l_1g\sin(\theta_1)=0
\label{8}
\end{equation}
and for $\theta_2$, we get \begin{equation}
m_2l_2^2\ddot{\theta_2}+m_2l_1l_2\ddot{\theta_1}\cos(\theta_1-\theta_2)-m_2l_1l_2\dot{\theta_1}^2\sin(\theta_1-\theta_2)+m_2l_2g\sin(\theta_2)=0
\label{9}
\end{equation}
The above formulas are the equations of motion of double pendulum \cite{la}.\\
Now, we will start to write Lagrangian of triple pendulum shown in \ref{fig:3} , and then, we will obtain the equation of motion of triple pendulum. These equations for simple pendulum, double pendulum, and triple pendulum will guide us to derive our final equation.\\
Firstly, \begin{equation}
x_1=l_1\sin(\theta_1) \quad y_1=-l_1\cos(\theta_1)
\label{10}
\end{equation}
\begin{equation}
x_2=l_1\sin(\theta_1)+ l_2\sin(\theta_2)\quad y_2=-(l_1\cos(\theta_1)+ l_2\cos(\theta_2))
\label{11}
\end{equation}
\begin{equation}
x_3=l_1\sin(\theta_1)+ l_2\sin(\theta_2)+ l_3\sin(\theta_3)\quad y_3=-(l_1\cos(\theta_1)+ l_2\cos(\theta_2)+l_3\cos(\theta_3))
\label{12}
\end{equation}
From \ref{2} and \ref{7}, we know the Lagrangian of simple and double pendulum.\\
Then,\begin{equation}
 dx_3= \cos(\theta_1)l_1d\theta_1+\cos(\theta_2)l_2d\theta_2+\cos(\theta_3)l_3d\theta_3
\end{equation} \begin{equation}
dy_3= \sin(\theta_1)l_1d\theta_1+\sin(\theta_2)l_2d\theta_2+\sin(\theta_3)l_3d\theta_3
\end{equation} 
Thus, 
\begin{equation}
\begin{aligned}
(dx_3)^2+(dy_3)^2 &=l_1^2(d\theta_1)^2+l_2^2(d\theta_2)^2+l_3^2(d\theta_3)^2+2l_1l_2\cos(\theta_1-\theta_2)d\theta_1d\theta_2 \\
&+(\theta_1-\theta_3)d\theta_1d\theta_3+ 2l_2l_3\cos(\theta_2-\theta_3)d\theta_2d\theta_3
\end{aligned}
\label{13}
 \end{equation}
Consequently,\begin{equation}
\begin{aligned}
K_3 & =\frac{1}{2}m_3(l_1^2(\dot{\theta_1})^2+l_2^2(\dot{\theta_2})^2+l_3^2(\dot{\theta_3})^2+2l_1l_2\cos(\theta_1-\theta_2)\dot{\theta_1}\dot{\theta_2}+ \\
& 2l_1l_3\cos(\theta_1-\theta_3)\dot{\theta_1}\dot{\theta_3}+
 2l_2l_3\cos(\theta_2-\theta_3)\dot{\theta_2}\dot{\theta_3})
 \end{aligned}
\end{equation} \begin{equation}
U_3=-m_3g(l_1\cos(\theta_1)+ l_2\cos(\theta_2)+l_3\cos(\theta_3))
\end{equation}.\\
Then, we get \begin{equation}
L_3=K_3-U_3\footnote{Since equation is very long, we give it implicitly.}
\label{14}
 \end{equation}
The Lagrangian of triple pendulum can be written as $L=L_1+L_2+L_3$.\\
If we want to write equation of motion of triple pendulum, we will write it with respect to $\theta_1$, $\theta_2$ and $\theta_3$ by use of Euler-Lagrange equation $ \frac{d}{dt}(\frac{\partial L}{\partial\dot{q}})-\frac{\partial L}{\partial q}=0$.\linebreak
For $\theta_1$,
\begin{equation}
\begin{aligned}
&gl_1(m_1\sin(\theta_1)+m_2\sin(\theta_1)+m_3\sin(\theta_1))+m_2l_1l_2\sin(\theta_1-\theta_2)\dot{\theta_1}\dot{\theta_2} \\
&+m_3l_1l_3\sin(\theta_1-\theta_3)\dot{\theta_1}\dot{\theta_3}  
+ m_3l_1l_2\sin(\theta_1-\theta_2)\dot{\theta_1}\dot{\theta_2} +l_1^2\ddot{\theta_1}(m_1+m_2+m_3) \\
&+m_2l_1l_2[\sin(\theta_2-\theta_1)(\dot{\theta_1}-\dot{\theta_2})\dot{\theta_2} +
 \cos(\theta_1-\theta_2)\ddot{\theta_2}] \\
 &+m_3l_1l_2[sin(\theta_2-\theta_1)(\dot{\theta_1}-\dot{\theta_2})\dot{\theta_2} +
 \cos(\theta_1-\theta_2)\ddot{\theta_2}] \\
 &+m_3l_1l_3[\sin(\theta_3-\theta_1)(\dot{\theta_1}-\dot{\theta_3})\dot{\theta_3}+\cos(\theta_1-\theta_3)\ddot{\theta_3}]=0
\end{aligned}
\label{15}
\end{equation}
For $\theta_2$,
 \begin{equation}
 \begin{aligned}
 & gl_2(m_2\sin(\theta_2)+m_3\sin(\theta_2))+\dot{\theta_1}\dot{\theta_2}l_1l_2\sin(\theta_2-\theta_1)[m_2+m_3] \\
 &+m_3l_2l_3\sin(\theta_2-\theta_3)\dot{\theta_2}\dot{\theta_3} +l_2^2\ddot{\theta_2}(m_2+m_3)\\  &+(m_2+m_3)l_1l_2[\sin(\theta_2-\theta_1)(\dot{\theta_1}-\dot{\theta_2})\dot{\theta_1}+\cos(\theta_2-\theta_1)\ddot{\theta_1}]\\
 &+m_3l_2l_3[\sin(\theta_3-\theta_2)(\dot{\theta_2}-\dot{\theta_3})\dot{\theta_3}+\cos(\theta_2-\theta_3)\ddot{\theta_3}]=0
 \end{aligned}
 \label{16}
  \end{equation}
For $\theta_3$,
\begin{equation}
\begin{aligned}
& m_3gl_3\sin(\theta_3)-m_3l_2l_3\sin(\theta_2-\theta_3)\dot{\theta_2}\dot{\theta_3}-m_3l_1l_3\sin(\theta_1-\theta_3)\dot{\theta_1}\dot{\theta_3}\\
&+m_3l_1l_3[\sin(\theta_3-\theta_1)(\dot{\theta_1}-\dot{\theta_3})\dot{\theta_1} 
 +\cos(\theta_1-\theta_3)\ddot{\theta_1}]\\
 &+m_3l_2l_3[\sin(\theta_3-\theta_2)(\dot{\theta_2}-\dot{\theta_3})\dot{\theta_2}+cos(\theta_2-\theta_3)\ddot{\theta_2}]+m_3l_3^2\ddot{\theta_3}=0
\end{aligned}
\label{17}
\end{equation}
Now, it can be seen that the terms in the equation of motion of simple, double and triple pendulum may be grouped. Also, if we can understand the behaviour of the terms which consists of cosine and sine functions, we can predict what kind of terms will occur in the equation of motion of pendulums that contains more than three point masses. Now, we will analyse the terms in the equation of motion of triple pendulum. 
Firstly, the term $gl_jsin(\theta_j)m_k$ is common in all the equations. However, the mass varies in three equation. \ref{15} has $m_1$, $m_2$, $m_3$. But when we look at \ref{16}, we lost $m_1$,and when we look at \ref{17}, we lost $m_2$. Let $\theta_j$ indicates the coordinates and $j=1,2,3$. Then by defining a function, call $\sigma_{jk}$, we can create this pattern in a sum.
$$\sigma_{jk}= \left\{
        \begin{array}{ll}
            0 & \quad j>k \\
            1 & \quad j\leq k
        \end{array}
    \right.$$
Therefore, \begin{equation}
\label{eq:18}
\sum_{k=1}^{n=3} gl_j\sin(\theta_j)m_k\sigma_{jk}
\end{equation} can give us the required terms when n=3 for triple pendulum.
Moreover, $m_3l_3^2\ddot{\theta_3}$ appears in three equation with same trend as we discussed for \ref{eq:18}. Thus, we can formulate it in the same manner by again using $\sigma_{jk}$.\\
\begin{equation}
\sum_{k=1}^{n=3} m_kl_j^2\ddot{\theta_j}\sigma_{jk}.
\label{19}
\end{equation}
In fact, we are now left with 2 different kind of terms. The first is $m_3l_1l_3\sin(\theta_1-\theta_3)\dot{\theta_1}\dot{\theta_3}$ appearing in \ref{17}. It can be seen that it consists of a $\theta_{j_1}$ and its combinations with $\theta_{j_2}$ and $\theta_{j_3}$(the same $\theta_{j_t}$ do not appear twice) from \ref{15},\ref{16},\ref{17}. \\
Now by following the trend we can make a formulation. It is 
\begin{equation}
\sum_{k=1}^{n=3} (\sum_{q\geq k}^{n=3} m_q\sigma_{jq})l_jl_k\sin(\theta_j-\theta_k)\dot{\theta_j}\dot{\theta_k}
\label{20}
\end{equation}
We dealt with minus signs by changing the order of arguments of sine function.\\
Also, the last terms is the one $m_3l_2l_3[\sin(\theta_3-\theta_2)(\dot{\theta_2}-\dot{\theta_3})\dot{\theta_2}+\cos(\theta_2-\theta_3)\ddot{\theta_2}]$. The way they appear in \ref{15},\ref{16},\ref{17} and general trend is similar.Thus, we can use a similar manner to derive o formula. Nevertheless, we need to define a new function, $\phi_{jk}$, to prevent the case when $cos(0)$ in our formulation. The new function is \\
$$\phi_{jk}= \left\{
        \begin{array}{ll}
            0 & \quad j=k \\
            1 & \quad j\neq k
        \end{array}
    \right.$$
Then the formulation is \\
\begin{equation}
\sum_{k=1}^{n=3} (\sum_{q\geq k}^{n=3} m_q\sigma_{jq})l_jl_k[\sin(\theta_k-\theta_j)[\dot{\theta_j}-\dot{\theta_k}]\dot{\theta_k}+\phi_{jk}\cos(\theta_j-\theta_k)\ddot{\theta_k}]
\label{21}
\end{equation}
Now, we will give the final form of our formulation for triple pendulum (n=3), and then, we will do the discussion for generalization of this formulation. \\
The final form is \\
\begin{equation}
\begin{aligned}
& \sum_{k=1}^{n=3} \Big(gl_j\sin(\theta_j)m_k\sigma_{jk}+m_kl_j^2\ddot{\theta_j}\sigma_{jk}+(\sum_{q\geq k}^{n=3} m_q\sigma_{jq})l_jl_k\sin(\theta_j-\theta_k)\dot{\theta_j}\dot{\theta_k} \\
& + (\sum_{q\geq k}^{n=3} m_q\sigma_{jq})l_jl_k[\sin(\theta_k-\theta_j)[\dot{\theta_j}-\dot{\theta_k}]\dot{\theta_k}+\phi_{jk}\cos(\theta_j-\theta_k)\ddot{\theta_k}]\Big) =0 
\end{aligned}
\label{22}
\end{equation} 
Firstly, we can generalize \ref{eq:18} and \ref{19} directly to n. The term in \ref{eq:18} is obtained from the partial derivative of Lagrangian with respect to the $\theta_j$ ,and its anti-derivative comes from the potential. Thus, it will appear in the same form as we formulated in eq.18. Moreover, the term in \ref{19} is obtained from the partial derivative of Lagrangian with respect to the $\dot{\theta_j}$, and total derivative with respect to time. Also its anti-derivative comes from the kinetic energy expression.Therefore, its trend of appearing in equation of motion for higher point masses will be the same with the \ref{19}. To understand how \ref{20} will behave for an arbitrary n, we need first to consider where the cosine subtraction form comes.
For an arbitrary n , $(dx_n)=\sum_{i=1}^{n}l_i\cos(\theta_i)d\theta_i$ and $(dy_n)=\sum_{i=1}^{n}l_isin(\theta_i)d\theta_i$.Thus, $(dx_n)^2+(dy_n)^2$ consists of only $l_i^2d\theta_i^2$ and the combination of $2l_kl_m\cos(\theta_k-\theta_m)d\theta_kd\theta_m$,where $1\leq m<k \leq n$. We obtain the term in \ref{20} from the partial derivative of Lagrangian with respect to the $\theta_j$, and we use $\sum_{q\geq k}^{n=3} m_q\sigma_{jq}$ to arrange masses. Consequently, we can directly generalize n to an arbitrary n where $n>0$. The last terms represented in \ref{21} is also comes from the same anti-derivative, but it is obtained from the partial derivative of Lagrangian with respect to the $\dot{\theta_j}$, and total derivative with respect to time.
Because of the same reasons, we can generalize it to an arbitrary n. A crucial point is whether there will be extra new terms for higher n's. Actually, we grouped our terms in four ,and we know where they come. When we generalize these terms to n, the resulting Lagrangian will be in the same form. Therefore, we will see the same types of terms when we put our Lagrangian into Euler-Lagrange Equation. Because of these reasons, we can safely say that there will be no new terms for higher n values. The equation is 
\begin{equation}
\begin{aligned}
& \sum_{k=1}^{n} \Big(gl_j\sin(\theta_j)m_k\sigma_{jk}+m_kl_j^2\ddot{\theta_j}\sigma_{jk}+(\sum_{q\geq k}^{n} m_q\sigma_{jq})l_jl_k\sin(\theta_j-\theta_k)\dot{\theta_j}\dot{\theta_k} \\
& + (\sum_{q\geq k}^{n} m_q\sigma_{jq})l_jl_k[\sin(\theta_k-\theta_j)[\dot{\theta_j}-\dot{\theta_k}]\dot{\theta_k}+\phi_{jk}cos(\theta_j-\theta_k)\ddot{\theta_k}]\Big) =0 
\end{aligned}
\label{23}
\end{equation}  

\section{Second Way of Derivation\protect\footnote{I would like to thank Professor Altug Ozpineci for his great contributions to this section.}}
\begin{figure}[]
\centering 
	\includegraphics[scale=0.45]{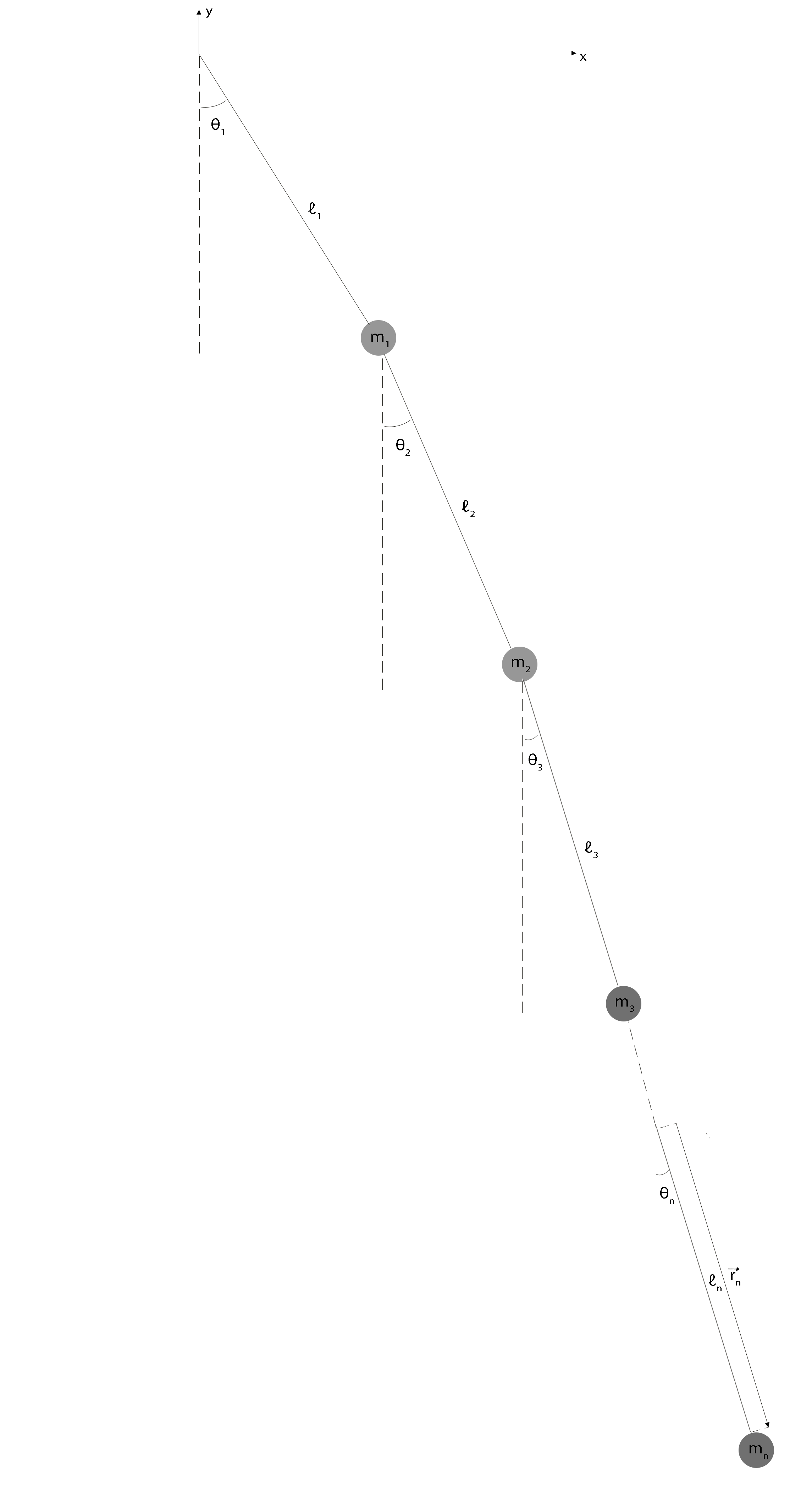}
	\caption{A Pendulum system with n point masses}
	\label{fig:4}
\end{figure}

In this section, we will try to obtain the same equation by using another method. Assume that we have a pendulum as illustrated in \ref{fig:4} and n point masses. Let us define a vector $\vec{r_n}=l_n\sin(\theta_n)\hat{x}-l_n\cos(\theta_n)\hat{y}$. Then
\begin{equation}
\vec{R_n}=\sum_{i=1}^{n} \vec{r_n}
\label{24}
\end{equation}
\begin{equation}
\dot{\vec{R_n}}=\sum_{i=1}^{n} l_i\dot{\theta_i}(\cos\theta_i\hat{x}+\sin\theta_i\hat{y})
\label{25}
\end{equation}
To be able to write kinetic energy expression, we need $|\dot{\vec{R_n}}|^2$. Thus, 
\begin{equation}
|{\dot{\vec{R_n}}}|^{2}=\vec{R_n}\cdot\vec{R_n}=\sum_{i,j=1}^{n} l_il_j\dot{\theta_i}\dot{\theta_j}\cos(\theta_i-\theta_j)
\label{26}
\end{equation}
Consequently, the kinetic energy and potential expressions are 
\begin{equation}
T=\sum_{k=1}^{n} \frac{1}{2}m_k|{\dot{\vec{R_n}}}|^{2}
\label{27}
\end{equation}
\begin{equation}
U=\sum_{i=1}^{n} U_i = -g \sum_{i=1}^{n} m_i \sum_{j=1}^{i} l_j\cos\theta_j
\label{28}
\end{equation}
Then, the Lagrangian of the system is 
\begin{equation}
L=T-U=\sum_{k=1}^{n} \frac{1}{2}m_k|{\dot{\vec{R_n}}}|^{2}+g \sum_{i=1}^{n} m_i \sum_{j=1}^{i} l_j\cos\theta_j
\label{29}
\end{equation}
Now, let us find equations of motion for an arbitrary $\theta_q$ $(q\leq n)$.
\begin{equation}
\frac{d}{dt}(\frac{\partial L}{\partial{\dot{\theta_q}}})-\frac{\partial L}{\partial{\theta_q}}=
-\frac{\partial T}{\partial{\theta_q}}+\frac{\partial(U)}{\partial{\theta_q}}+\frac{d}{dt}(\frac{\partial T}{\partial{\dot{\theta_q}}})=0
\label{46}
\end{equation}
Thus, we have 
\begin{equation}
\frac{\partial(U)}{\partial{\theta_q}}=g \sum_{i=q}^{n} m_il_q\sin(\theta_q)
\tag{46.1} \label{46a}
\end{equation}
\begin{equation}
\begin{aligned}
-\frac{\partial T}{\partial{\theta_q}} & =-\frac{\partial}{\partial{\theta_q}}\Big(\sum_{k=1}^{n} \frac{1}{2}m_k\sum_{i,j=1}^{k} l_il_j\dot{\theta_i}\dot{\theta_j}cos(\theta_i-\theta_j) \Big) \\
 & =\sum_{k=q}^{n} m_k\sum_{i=1}^{k} l_il_q\dot{\theta_i}\dot{\theta_q}\sin(\theta_q-\theta_i)
 \end{aligned}
\tag{46.2} \label{46b}
\end{equation}
Now, we have two special case in kinetic energy expression \ref{26}. The first one is when $i=j$ , and this one gives terms in the form of $\frac{1}{2}mv^2$,which contains $\dot{\theta_i}^2$ terms, and the second case is when $i \neq j$.Thus, we will investigate $\frac{d}{dt}(\frac{\partial T}{\partial{\dot{\theta_q}}})$ in two different parts. \\
When $i=j$, 
\begin{equation}
\begin{aligned}
& \frac{d}{dt}\big(\frac{\partial }{\partial{\dot{\theta_q}}}\Big(\sum_{k=1}^{n} \frac{1}{2}m_k\sum_{i=j=1}^{k} l_il_j\dot{\theta_i}\dot{\theta_j}\cos(\theta_i-\theta_j) \Big)\big) \\
& = \sum_{k=q}^{n} m_k{l_q}^2\ddot{\theta_q} 
\end{aligned}
\tag{46.3} \label{46c}
\end{equation}
When $i \neq j$,
\begin{equation}
\begin{aligned}
& \frac{d}{dt}\big(\frac{\partial }{\partial{\dot{\theta_q}}}\Big(\sum_{k=1}^{n} \frac{1}{2}m_k\sum_{i,j=1}^{k} l_il_j\dot{\theta_i}\dot{\theta_j}\cos(\theta_i-\theta_j) \Big)\big) \\
& = \sum_{k=q}^{n} m_k\sum_{i=1}^{k} l_il_q\dot{\theta_i}\dot{\theta_q}\sin(\theta_i-\theta_q)[\dot{\theta_q}-\dot{\theta_i}]  \\
& + \sum_{k=q}^{n} m_k\sum_{i=1}^{k} l_il_q\cos(\theta_i-\theta_q)\ddot{\theta_i}
\end{aligned}
\tag{46.4} \label{46d}
\end{equation}
Since we indicate that $i \neq j$, the cases $\cos(0)$ and $\sin(0)$ are automatically prevented in \ref{46d}. Nevertheless, we cannot prevent $\cos(0)$ case when $q=i$. To solve this problem, we need to define $\phi_{iq}$ function, which we defined in the previous section.
When we group \ref{46a},\ref{46b},\ref{46c} and \ref{46d}, we acquire the equations of motion for an arbitrary $\theta_q$ 
\begin{equation}
\begin{aligned}
& g\sum_{i=q}^{n} m_il_q\sin(\theta_q) + \sum_{k=q}^{n} m_k\sum_{i=1}^{k} l_il_q\dot{\theta_i}\dot{\theta_q}\sin(\theta_q-\theta_i) +\sum_{k=q}^{n} m_k{l_q}^2\ddot{\theta_q} \\
& +\sum_{k=q}^{n} m_k\sum_{i=1}^{k} l_il_q\dot{\theta_i}\dot{\theta_q}sin(\theta_i-\theta_q)[\dot{\theta_q}-\dot{\theta_i}]  
 + \sum_{k=q}^{n} m_k\sum_{i=1}^{k} l_il_q\phi_{iq}\cos(\theta_i-\theta_q)\ddot{\theta_i}=0
\end{aligned}
\label{31}
\end{equation}
When \ref{31} is compared with \ref{23}, they are exactly the same except summation form. This is because we use a function called $\sigma_{jk}$ and $\phi_{jk}$ to write the formula in a compact one big summation; however, we have double sums in \ref{31}. Also, \ref{31} can be written in the parenthesis of a general sum. Since $i$ is a dummy index in \ref{46a}, replace it with $k$. Now it can be written in the form of 
\begin{equation}
\begin{aligned}
\sum_{k=q}^{n} & \Big( gm_kl_q\sin(\theta_q) + m_k\sum_{i=1}^{k} l_il_q\dot{\theta_i}\dot{\theta_q}\sin(\theta_q-\theta_i) + m_k{l_q}^2\ddot{\theta_q} \\
& + m_k\sum_{i=1}^{k} l_il_q\dot{\theta_i}\dot{\theta_q}\sin(\theta_i-\theta_q)[\dot{\theta_q}-\dot{\theta_i}] + m_k\sum_{i=1}^{k} l_il_q\phi_{iq}\cos(\theta_i-\theta_q)\ddot{\theta_i} \Big)=0
\end{aligned}
\label{32}
\end{equation}
\pagebreak
\section{Investigation of Small Oscillations}
In this section, we will basically consider that given $\theta_j$'s are sufficiently small to do small angle approximation, and try to see whether it is matching with the small angle approximation used for simple, double and other pendulum systems. That is, we will assume the followings
\begin{equation}
\sin(\theta)\approx \theta ,
\end{equation}
\begin{equation}
\cos(\theta)\approx 1-\frac{\theta^2}{2}
\end{equation}
Then,
\begin{equation}
\begin{aligned}
& \sum_{k=1}^{n}\Big(gl_j\theta_jm_k\sigma_{jk}+m_kl_{j}^2\ddot{\theta_j}\sigma_{jk}+(\sum_{q \geq k}^{n} m_q\sigma_{jq})l_jl_k(\theta_j-\theta_k)\dot{\theta_j}\dot{\theta_k}+\\
&(\sum_{q \geq k}^{n}  m_q\sigma_{jq})l_jl_k[(\theta_k-\theta_j)(\dot{\theta_j}-\dot{\theta_k})\dot{\theta_k}+\phi_{jk}(1-\frac{(\theta_j-\theta_k)^2}{2})\ddot{\theta_k}]\Big) \\
&=\sum_{k=1}^{n} \Big(\theta_j(gl_jm_k\sigma_{jk})+m_kl_{j}^2\ddot{\theta_j}\sigma_{jk}+\theta_j(\sum_{q \geq k}^{n}  m_q\sigma_{jq})l_jl_k\dot{\theta_k}^2+
\\
&(\sum_{q \geq k}^{n}  m_q\sigma_{jq})l_jl_k[-\theta_k\dot{\theta_k}^2+\phi_{jk}\ddot{\theta_k}-\phi_{jk}\ddot{\theta_k}\frac{\theta_{j}^2}{2}-\phi_{jk}\ddot{\theta_k}\frac{\theta_{k}^2}{2}+\theta_k\theta_j\phi_{jk}\ddot{\theta_k}]\Big)=0
\end{aligned}
\end{equation}
We have
\begin{equation}
\begin{aligned}
&\sum_{k=1}^{n} \Big(\theta_j(gl_jm_k\sigma_{jk})+m_kl_{j}^2\ddot{\theta_j}\sigma_{jk}+(\sum_{q \geq k}^{n}  m_q\sigma_{jq})l_jl_k\theta_j\dot{\theta_k}^2+
\\
&(\sum_{q \geq k}^{n}  m_q\sigma_{jq})l_jl_k[-\theta_k\dot{\theta_k}^2+\phi_{jk}\ddot{\theta_k}-\phi_{jk}\ddot{\theta_k}\frac{\theta_{j}^2}{2}-\phi_{jk}\ddot{\theta_k}\frac{\theta_{k}^2}{2}+\theta_k\theta_j\phi_{jk}\ddot{\theta_k}]\Big)=0
\end{aligned}
\label{33}
\end{equation}
In the case of simple pendulum, we have $n=1$ and $j=1$. For $n=1$ and $j=1$, 
\begin{equation}
\begin{aligned}
&\theta_1gl_1m_1+m_1l_1^2\ddot{\theta_1}+\theta_1m_1l_1^2\dot{\theta_1}^2-\theta_1m_1l_1^2\dot{\theta_1}^2+0-0-0+0
\\
&=\theta_1gl_1m_1+m_1l_1^2\ddot{\theta_1}=0 \\
& \Rightarrow \ddot{\theta_1}+\frac{g}{l}\theta_1=0
\end{aligned}
\end{equation}
In the case of double pendulum, we have $n=2$ and $j={1,2}$. For $j=1$, 
\begin{equation}
\begin{aligned}
& gl_1\theta_1m_1+m_1l_1^2\ddot{\theta_1}+\theta_1(m_1+m_2)l_1^2\dot{\theta_1^2}-(m_1+m_2)\theta_1\dot{\theta_1}^2l_1^2+\theta_1gl_1m_2+\\
&m_2l_1^2\ddot{\theta_1}+\theta_1m_2l_1l_2\dot{\theta_2}^2-m_2l_1l_2\theta_2\dot{\theta_2}^2+(m_2)l_1l_2(\ddot{\theta_2}-\frac{\theta_1^2}{2}\ddot{\theta_2}-\frac{\theta_2^2}{2}\ddot{\theta_2}+\theta_1\theta_2\ddot{\theta_2})=0
\end{aligned}
\end{equation}
Therefore, it can be seen that the the results are matching and we successfully manage to obtain general version of small angle approximation for pendulum systems by using \ref{23}.

\section{An Analysis on Infinite Number of Point Masses}
\indent We will start with \ref{33}. First of all we want to analyze our n-point mass system under small oscillations and we know that at small angles we may represent our system as a system of linear equations. However, we have still non-linear terms in \ref{33}. Now consider the terms $\theta_j\dot{\theta_k}^2$ and $-\theta_k\dot{\theta_k}^2-\phi_{jk}\ddot{\theta_k}\frac{\theta_{j}^2}{2}-\phi_{jk}\ddot{\theta_k}\frac{\theta_{k}^2}{2}+\theta_k\theta_j\phi_{jk}\ddot{\theta_k}$ in \ref{33}. Since any $\theta_j$ is very small($\theta_j<< 1$), all the terms containing multiplication of several arbitrary $\theta_j$ and their derivatives can be neglected. Thus, the terms indicated above vanishes, and we are left with
\begin{equation}
\sum_{k=1}^{n} \Big(\theta_j(gl_jm_k\sigma_{jk})+m_kl_{j}^2\ddot{\theta_j}\sigma_{jk}+(\sum_{q \geq k}^{n}  m_q\sigma_{jq})l_jl_k\phi_{jk}\ddot{\theta_k}\Big)=0
\label{34}
\end{equation}
Now, we will set our assumptions.
\begin{itemize}
\item Assume that we have initially an n-point mass pendulum system.
\item Let $l_1=l_2=....=l_n=l$ and $m_1=m_2=....=m_n=m$ .
\item The total length of the system is $L=\sum_{i=1}^{n} l_i=nl$.($l$ is the length of the line segment which connects adjacent point masses.)
\item The total length of the system is $M=\sum_{i=1}^{n} m_i=nm$.
\item For any $\theta_j(1\leq j \leq n)$, $\theta_j<< 1$.
\item By fifth and second assumptions, $\theta_j=(x_j-x_{j-1})/l$
\end{itemize}
At this stage, the mass density(not continuous) of our system is
\begin{equation}
\mu=\frac{M}{L}
\end{equation}
Our aim is to take limit as $n \rightarrow \infty$. To do so, we need to re-arrange \ref{34} and 
construct a well-defined system. Thus, we will re-define some terms. \\
Our limit approach will be based on sequential concepts. We increase $n$ by inserting a new point mass between every two point mass, so each step the total number of point masses grows in power of two. By following these steps, $l$ becomes half of its previous value in each steps. Nevertheless, we have no control on mass of individual point masses. Consequently as  $n \rightarrow \infty$, $M \rightarrow \infty$. Thus, $\mu \rightarrow \infty$ ,which is very problematic situation. 

Therefore, we must do some sequential definitions. That is
\begin{itemize}
\item $(n_p)=n2^p$ ,where $p=0,1,2,3,...$ and $n$ is the initial number of point masses in our system.
\item $(l_p)={l}/{2^p}$, where $p=0,1,2,3,...$ and $l$ is the initial length of a line segment between point masses in our system.
\item $(m_p)={m}/{2^p}$, where $p=0,1,2,3,...$ and $m$ is the initial mass of a point mass in our system.
\end{itemize}
All of these definitions indicate that while we are increasing $n$ in each step, we must reduce $m$ and $l$ values to half of their values in previous step.\pagebreak \\
Now, one can observe that when we take limit of $\mu$ as $p \rightarrow \infty $(total number of point masses $\rightarrow \infty$), the mass density stays constant and since $l \rightarrow 0$ it has the same value in any segment of our system. However, we still do not know what kind of a system it will be. Before starting taking
limits, we must do necessary arrangements in our equation.
The first sum in \ref{34} can be written as 
\begin{equation}
\begin{aligned}
\sum_{k=1}^{n} \theta_jgl_jm_k\sigma_{jk}& =\sum_{k=1}^{n} glm\sigma_{jk}\frac{(x_j-x_{j-1})}{l} \\
& = (n-j+1)mlg\frac{(x_j-x_{j-1})}{l}
\end{aligned}
\label{35}
\end{equation}
The second sum in \ref{34} can be written as 
\begin{equation}
\begin{aligned}
\sum_{k=1}^{n} \ddot{\theta_j}l_j^2m_k\sigma_{jk}& =\sum_{k=1}^{n} l^2m\sigma_{jk}\frac{(\ddot{x_j}-\ddot{x}_{j-1})}{l} \\
& = (n-j+1)ml^2\frac{(\ddot{x_j}-\ddot{x}_{j-1})}{l}
\end{aligned}
\label{36}
\end{equation}
For the third term we need to investigate summation in two cases because of the natures $\phi$ and $\sigma$ functions. The first case is when $k<j$. It may be written as 
\begin{equation}
\sum_{k=1}^{j-1} (\sum_{q \geq k}^{n}  m_q\sigma_{jq})l_jl_k\phi_{jk}\ddot{\theta_k}=\sum_{k=1}^{j-1} ml^2(n-j+1)\frac{(\ddot{x_k}-\ddot{x}_{k-1})}{l}
\end{equation}
For $k=1$, we have $x_0=0$. Hence, the sum is \\
\begin{equation}
\sum_{k=1}^{j-1} (\sum_{q \geq k}^{n}  m_q\sigma_{jq})l_jl_k\phi_{jk}\ddot{\theta_k}=(n-j+1)ml^2\frac{\ddot{x}_{j-1}}{l}
\label{37}
\end{equation}
When $k>j$,
\begin{equation}
\begin{aligned}
& \sum_{k=j+1}^{n} (\sum_{q \geq k}^{n}  m_q\sigma_{jq})l_jl_k\phi_{jk}\ddot{\theta_k} =\frac{ml^2}{l}((n-j)(\ddot{x}_{j+1}-\ddot{x}_{j})+(n-j-1)(\ddot{x}_{j+2}-\ddot{x}_{j+1})\\
& +\sum_{k=j+3}^{n} (n-k+1)(\ddot{x}_k-\ddot{x}_{k-1}))= \frac{ml^2}{l}(-(n-j)\ddot{x}_j+\ddot{x}_{j+1} \\
&+(n-j-1)\ddot{x}_{j+2}+\sum_{k=j+3}^{n} (n-k+1)(\ddot{x}_k-\ddot{x}_{k-1}))
\end{aligned}
\end{equation}
As one can observe, there will be another $\ddot{x}_{j+2}$ term when we expand the summation one more step and that sum will reduce the coefficient of $\ddot{x}_{j+2}$ to 1. After some algebra, the coefficient of $\ddot{\theta}_k$'s will be 1. Therefore, our summation becomes equal to very simple expression\\
\begin{equation}
\sum_{k=j+1}^{n} (\sum_{q \geq k}^{n}  m_q\sigma_{jq})l_jl_k\phi_{jk}\ddot{\theta_k} =
\frac{ml^2}{l}(-(n-j)\ddot{x}_j+\sum_{k=j+1}^{n} \ddot{x}_k)
\label{38}
\end{equation}
Now, let us substitute \ref{35}, \ref{36}, \ref{37} and \ref{38} into \ref{34}. We have\\
\begin{equation}
\begin{aligned}
& (n-j+1)mlg\frac{(x_j-x_{j-1})}{l} + (n-j+1)ml^2\frac{(\ddot{x_j}-\ddot{x}_{j-1})}{l}\\
& +(n-j+1)ml^2\frac{\ddot{x}_{j-1}}{l}+\frac{ml^2}{l}(-(n-j)\ddot{x}_j+\sum_{k=j+1}^{n}\ddot{x}_k)s=0
\end{aligned}
\end{equation}
After cancellations,\\
\begin{equation}
\begin{aligned}
& (n-j+1)mlg\frac{(x_j-x_{j-1})}{l}+\frac{ml^2}{l}\ddot{x}_j+\frac{ml^2}{l}\sum_{k=j+1}^{n}\ddot{x}_k=0
\end{aligned}
\label{39}
\end{equation}
Observe that\\
 \begin{equation}
\begin{aligned}
& \frac{ml^2}{l}\sum_{k=j}^{n}\ddot{x}_k=-(n-j+1)mlg\frac{(x_j-x_{j-1})}{l}
\end{aligned}
\label{40}
\end{equation}
Let us substitute \ref{40} into \ref{39}. Since $j$ is dummy index in \ref{40}, let $j=j+1$ to obtain correct indices after substitution. Moreover, divide \ref{39} with $ml$. Consequently, we have\\
\begin{equation}
\begin{aligned}
\ddot{x}_j-(n-j)g\frac{(x_{j+1}-x_{j})}{l}+(n-j+1)g\frac{(x_j-x_{j-1})}{l}=0
\end{aligned}
\end{equation}
Also, we can say that $x_j$ may be expressed as a function of $y$ and $t$. Let us substitute $n$ and $l$ in their sequential forms. Then, after some algebra, we have\\
\begin{equation}
\begin{aligned}
\ddot{x}_j=g[((n_p)-j)\frac{(x_{j+1}-2x_{j}+x_{j-1})}{(l_p)}-\frac{(x_j-x_{j-1})}{(l_p)}]
\end{aligned}
\label{41}
\end{equation} 
Let us start taking limit as $p \rightarrow \infty$. Firstly,
\begin{equation}
\lim_{p \rightarrow \infty}	\ddot{x_j}=\frac{\partial^2 x(y,t)}{\partial t^2}
\end{equation}
As $p \rightarrow \infty$, $(n_p) \rightarrow \infty$. 	Thus, our system becomes a continuous system since 
\begin{equation}
\lim_{p \rightarrow \infty} (x_{j+1}-x_{j})=0
\end{equation}
Consequently,  $\ddot{x_j}$ is an arbitrary point on the graph of $\partial^2 x(y,t)/ \partial t^2$ and above limit is valid. \\
Now, consider the following 
\begin{equation}
\lim_{p \rightarrow \infty} \frac{(x_{j}-x_{j-1})}{(l_p)}=\lim_{p \rightarrow \infty} \frac{(x_{j}-x_{j-1})}{(l/2^p)}
\end{equation}
Since $x_j$ and $x_{j-1}$ differ by $l/2^p$ and $l/2^p$,say $h$, goes zero as $p$ goes infinity. Above limit is nothing but limit definition of first partial derivative of $x(y,t)$ with respect to $y$ at any point in the domain of  $x(y,t)$ since $j$ is arbitrary.\\
Ergo,
\begin{equation}
\lim_{p \rightarrow \infty} \frac{(x_{j}-x_{j-1})}{(l/2^p)}=\frac{\partial x(y,t)}{\partial y}
\label{42}
\end{equation}
Moreover, observe that 
\begin{equation}
\lim_{p \rightarrow \infty} ((n_p)(l_p)-(l_p)j)=L-y
\end{equation}
We must add $y$ term since $(l_p)j$ will give us y-coordinate value of $j^{th}$ point. There will be infinitely many points on our system and $(l_p)$ will be infinitely small.
Hence,
\begin{equation}
\lim_{p \rightarrow \infty} ((n_p)-j)\frac{(x_{j+1}-2x_{j}+x_{j-1})}{(l_p)}=\lim_{p \rightarrow \infty}((n_p)-j)\frac{(l_p)}{(l_p)}\frac{(x_{j+1}-2x_{j}+x_{j-1})}{(l_p)}
\end{equation}
By following the same manner we followed for \ref{42}, we will obtain
\begin{equation}
\lim_{p \rightarrow \infty}((n_p)-j)\frac{(l_p)}{(l_p)}\frac{(x_{j+1}-2x_{j}+x_{j-1})}{(l_p)}=(L-y)\frac{\partial^2 x(y,t)}{\partial y^2}
\end{equation}
Finally, we retrieve the following by taking limit of both sides in \ref{41}.
\begin{equation}
\frac{\partial^2 x(y,t)}{\partial t^2}=g(L-y)\frac{\partial^2 x(y,t)}{\partial y^2}-g\frac{\partial x(y,t)}{\partial y}
\label{43}
\end{equation}
As one can conclude, we have obtained a wave equation. However, it is in an unusual form. At the beginning of this section, we had no idea what we will see after the calculations. Fortunately, our limits happen to exist, and we have a reliable outcome. \\
Further analysis of \ref{43} can be done by using separation of variables method. After the calculations, we will have two ordinary differential equations. Now, we will demonstrate those calculations; however, we will not investigate the solutions since we are not interested in what type of solutions that our equations yield in this paper as our motive is to create a general and analytical analysis on this subject.\\
But first, let us investigate what we have found in \ref{43}.  After seeing the wave equation, we may have a candidate, and it may be the equation of motion of hanging rope with constant mass density. To confirm that we can use Euler-Lagrange Equation for continuous systems \cite{go} in the case of hanging rope with constant mass density. In this part, we will use our variables and notations to be able to compare the results. The Euler-Lagrange Equation is 
\begin{equation}
\frac{d}{dt}\frac{\partial \mathcal{L}}{\partial \frac{\partial x}{\partial t}}+
\frac{d}{dy}\frac{\partial \mathcal{L}}{\partial \frac{\partial x}{\partial y}}-\frac{\partial \mathcal{L}}{\partial x}=0
\end{equation}
, where $\mathcal{L}$ is our Lagrange Density. The Lagrangian is in the form of
\begin{equation}
L= \int dT-\int dU
\label{44}
\end{equation}
The potential energy of an infinitesimal piece of rope, say $dU$, might be found by using tension force due to the rope part below our infinitesimal piece. Thus, it can be written as
\begin{equation}
\begin{aligned}
& dl=(1+\frac{1}{2}(\frac{\partial x}{\partial y})^2)dy \\
& dh=\frac{1}{2}(\frac{\partial x}{\partial y})^2dy \\
& \Rightarrow dU=m_{below}gdh=g(L-y)\mu \frac{1}{2}(\frac{\partial x}{\partial y})^2dy
\end{aligned}
\end{equation}
Similarly, the kinetic energy of that infinitesimal piece is 
\begin{equation}
dT=\frac{1}{2}mv^2=\frac{1}{2}\mu dl (\frac{\partial x}{\partial t})^2
\end{equation}
Since we are interested in oscillations with small angles, $dl \rightarrow dy$. Consequently, 
\begin{equation}
dT=\frac{1}{2}mv^2=\frac{1}{2}\mu dy (\frac{\partial x}{\partial t})^2
\end{equation}
After substituting $dT$ and $dU$ in \ref{44} and setting integral boundaries to 0 and L, we have
\begin{equation}
\frac{1}{2} \int_{0}^{L}\big(\mu (\frac{\partial x}{\partial t})^2-g(L-y)\mu (\frac{\partial x}{\partial y})^2  \big)dy
\end{equation}
The integrand is the Lagrange Density $\mathcal{L}$. We will skip the part of taking this integral and other algebraic arrangements and give the result directly. It is
\begin{equation}
\frac{\partial^2 x(y,t)}{\partial t^2}=g(L-y)\frac{\partial^2 x(y,t)}{\partial y^2}-g\frac{\partial x(y,t)}{\partial y}
\label{45}
\end{equation}
This is exactly the same equation with \ref{43}! Consequently, this proves our arguments and concludes that when we have an n-point mass pendulum system with well-defined properties, the limit of this system as number of point masses goes infinity will yield the equation of motion of a hanging rope with constant mass density. Thus, our system will start moving exactly as a hanging rope, and this ends our analysis on this section.\\
We have finished our analysis, but let us present more information concerning \ref{43}. Firstly, we will attack by using separation of variables. Thus, let
\begin{equation}
x(y,t)=\Psi(y)\xi(t)
\end{equation}
After substituting it into equation and diving both sides by $\Psi(y)\xi(t)$, we get 
\begin{equation}
\frac{1}{\xi}\frac{d^2\xi}{dt^2}=\frac{g}{\Psi}\Big((L-y)\frac{d^2 \Psi}{dy^2}-\frac{d\Psi}{dy}\Big)=-K
\end{equation}
Thus, we have two ordinary differential equations. The first one is
\begin{equation}
\frac{d^2\xi}{dt^2}+K\xi=0
\end{equation}
,which is very familiar equation. The second one is 
\begin{equation}
(L-y)\frac{d^2 \Psi}{dy^2}-\frac{d\Psi}{dy}+\frac{K}{g}\Psi=0
\end{equation}
We will not go into details, but after using $h^2=\frac{1}{g}(L-y)$ substitution, one obtains the zeroth order Bessel's Equation. These have crucial importance for solutions of this system.\pagebreak \\
As we have pointed out before, this wave equation is in an unfamiliar form. At first glance, it may look strange; however, it is not. Firstly, origin of the term $g\frac{\partial x(y,t)}{\partial y}$ in \ref{43} can be traced back. In fact, it comes from our gravitational potential term. This implies whenever our system is in gravitational potential, this term must appear. Moreover, terms in the right-hand side of \ref{43} are multiplied by $g$. This indicates that when there is no gravitational potential, the system will stay in its initial state since there will be no restoring force or tension. Also, it is not going to move unless it has an initial velocity. These all are matching with the \ref{43}. Now, all those strange terms can be understood as gravitational effects on our systems. Similarly, similar terms appear in the hanging chain problem.
\section{Conclusion}
In this work, it is shown that the equations of motion of a pendulum containing n point masses can be formulated for an arbitrary value of n $(n>0)$ by using two different methods. The equation for an arbitrary $\theta_j$ $(j\leq n)$ is
\begin{equation}
\begin{aligned}
& \sum_{k=1}^{n} \Big(gl_j\sin(\theta_j)m_k\sigma_{jk}+m_kl_j^2\ddot{\theta_j}\sigma_{jk}+(\sum_{q\geq k}^{n} m_q\sigma_{jq})l_jl_k\sin(\theta_j-\theta_k)\dot{\theta_j}\dot{\theta_k} \\
& + (\sum_{q\geq k}^{n} m_q\sigma_{jq})l_jl_k[\sin(\theta_k-\theta_j)[\dot{\theta_j}-\dot{\theta_k}]\dot{\theta_k}+\phi_{jk}\cos(\theta_j-\theta_k)\ddot{\theta_k}]\Big) =0 
\end{aligned}
\label{54}
\end{equation}  
Additionally, we have shown that by constructing well-defined assumptions and taking limit, our n-point mass pendulum system converges to a hanging rope.
\ack
I would like to thank Professor Altuğ Özpineci and Professor Bayram Tekin for their extraordinary support and advice.
Also, the author thanks to Ege Can Karanfil for his kind help with the figures, as well as Cihan Yeşil and Zeki Seskir for their encouragements and advice. 

{}

\end{document}